\documentstyle[preprint,aps]{revtex}

\begin{document}

% \draft command makes pacs numbers print
\draft

% force linebreaks with \\
\title{
Phase Transitions in the Multicomponent Widom--Rowlinson Model
and in Hard Cubes on the BCC--Lattice
}

% repeat the \author\address pair as needed
\author{
P. Nielaba 
}

\address{
Institut f\"ur Physik, Universit\"at Mainz. D-55099 Mainz, Germany
}

\author{
J.L. Lebowitz\footnote{{\bf Permanent Address}:  Departments of Mathematics and Physics, Rutgers
University, New Brunswick, NJ 08903, U.S.A.}
%~\cite{byline1}
}

\address{
I.H.E.S., 35, Route de Chartres, F-91440 Bures--Sur--Yvette, France
}

\date{\today}
\maketitle
\begin{abstract}

We use Monte Carlo techniques and analytical methods to study the phase
diagram of the $M$--component Widom--Rowlinson model 
 on the bcc--lattice: there are $M$ species all with the same fugacity $z$ and a nearest
neighbor hard core exclusion between unlike particles. Simulations show that
for $M \geq 3$ there is a ``crystal
phase'' for $z$ lying between $z_c(M)$ and $z_d(M)$ while for $z > z_d(M)$
there are $M$ demixed phases each consisting mostly of one species.
For $M=2$ there is a direct second order transition from the gas
phase to the demixed phase 
 while for  $M \geq 3$ 
the transition at $z_d(M)$ appears to be first order 
putting it in the Potts model universality class. 
For $M$ large, Pirogov-Sinai theory gives $z_d(M) \sim M-2+2/(3M^2) +
... $.  
In the crystal phase the particles preferentially occupy one of the
sublattices, independent of species, i.e.\ 
spatial symmetry but not particle symmetry is broken.
For $M \to \infty$ this transition 
approaches that of the one component hard cube gas with
fugacity $y = zM$. We find by direct simulations of such a system a
transition at $y_c \simeq 0.71$ which is consistent with the simulation
$z_c(M)$  for large
$M$.  This
transition  appears to be always of the Ising type.
\end{abstract}

%\pacs{PACS numbers: 64.60.Cn, 05.50.+q, 02.70.Lq, 75.10.Hk}

%\narrowtext
%\twocolumn
%\begin{multicols}{1}

%%%%%%%%%%%%%%%%%%%%%%%%%%%%%%%%%%%%%%%%%%%%%%%%%%%%%%%%%%%%%%%%%%%%%%%%%%%
%%%%%%%                       BODY OF TEXT
%%%%%%%%%%%%%%%%%%%%%%%%%%%%%%%%%%%%%%%%%%%%%%%%%%%%%%%%%%%%%%%%%%%%%%%%%%%

\newpage

\section{Introduction}

The Widom-- Rowlinson (WR) model, introduced in 1970~\cite{r1}
as an ingeniously simple model for the study of
phase transitions in continuum fluids 
(for an overview see Ref.~\cite{Sl}),
continues to be, like its
authors, a rich source of insights and analytical results in many 
(sometimes quite unexpected) areas~\cite{r3,r2,r4,r4a}
of statistical mechanics. In this paper, dedicated with great pleasure
to Ben Widom on the occasion of his seventieth birthday,  
we continue our study of a variation of the original model from two to $M$ 
components on a lattice: hard core exclusion between particles of
different species on nearest neighbor sites.

This model was first considered by Runnels and Lebowitz~\cite{r9}
who proved that when the number of components $M$ is larger than some
minimum $M_0$ then the transition from the gas phase at small values
of $z$ to the demixed phase at large values of $z$ does not take place
directly. Instead there is, at intermediate values of $z$, $z_c < z < z_d$, an
ordered phase in which one of the sublattices (even or odd) is
preferentially occupied, i.e.\ there is a crystalline
(antiferromagnetically ordered) phase in which the average particle
density on the even and odd sublattices, $\rho_{e}$ and $\rho_{o}$ are
unequal. The average density, $\rho(I)$, of species $I = 1,\dots,M$, on
each sublattice is the
same for each $I$, with  $\rho_e(I) = M^{-1} \rho_e$ and
$\rho_o(I)= M^{-1} \rho_o$. The nature of the symmetry breaking is thus
very different from that in the demixed phase at $z > z_d$ where $\rho_{e}
= \rho_{o} = \rho$ but there exists one species, say $I'$, for which
$\rho(I') > M^{-1}\rho$.  The origin of this crystalline phase 
is purely entropic.  For $z$ fixed and $M$
large ``it pays'' for the system ``entropy wise'' to occupy just
one sublattice without any constraint; since there are no interactions between
particles on the same sublattice there are $M$ independent choices at each
site if we keep one of the sublattices
empty.  This more than compensates, at some $M > M_0$, for the ``loss'' of
``fugacity energy'' occasioned by keeping down the density in one of the 
sublattices.

A natural question now arises, just how big does $M_0$ have to be to see
this ordered phase for $M \geq M_0$. It was shown in~\cite{r9}
that on the square lattice $M_0 < 27^6$; a ridiculously large upper
bound. On the other hand a direct computation on the Bethe 
lattice~\cite{LNS,LMNS} with $q$--neighbors gives $M_0 = [q/(q-2)]^2$,
which would suggest $M_0 \sim 4$ for the square lattice, $M_0 \sim 3$
for the cubic and $M_0 \sim 2$ for the bcc lattice. 
Now it can be shown, using
FKG inequalities, that $M_0 \geq 3$ on any bipartite lattice~\cite{x1},
but beyond that we have no simple or convincing argument for
finding $M_0$.
We therefore turned to 
Monte Carlo simulations.  This gave  on the square lattice 
$M_0 = 7$~\cite{LNS,LMNS} which is only about twice as large as the Bethe
lattice prediction.  This wetted our appetite to try the bcc lattice where
$q=8$.   
To our surprise we find here, using Monte Carlo simulations, that $M_0$ does
indeed equal $3$, on the bcc lattice. 

While we have no clue of how to find 
rigorously the actual value of $M_0$ or of $z_c(M)$,  it was argued in
~\cite{LMNS}  that 
for a given $z$ and $M$ large enough, the typical
occupancy pattern on the lattice (ignoring the label $I$ of the particles)
should be like that of a one component lattice gas with nearest neighbor hard
core exclusion.  For the latter system Dobrushin~\cite{r5} proved the existence of a
crystalline state.  
This implies that $z_c(M)$ should behave for large $M$ as $y_c/M$, where $y_c$ is
the critical fugacity at which the one component hard cube gas (occupation at 
a site $j$ excludes occupation at all eight neighboring sites)
crystalizes. The value of $y_c$ for the bcc lattice, obtained by
Gaunt~\cite{GAUNT} using series expansion methods, is $ 0.77 \pm 0.05$.
Using MC we obtained $y_c = 0.71 \pm 0.01$ which is roughly consistent with 
Gaunt's value. Our result also agrees well with the values of $M z_c(M)$ for
large $M$ being approximately $.72 \pm 02$ for $M=50$ and $M=100$.  This provides solid evidence for the existence of a reentry phase
transition in the temperature-- magnetic field plane phase diagram of an
Ising spin system with nearest-neighbor anti-ferromagnetic interactions on
the bcc lattice~\cite{DPL} .  

We also find, 
as in~\cite{LMNS}, that for large $M$,
$z_d(M)$  for the crystal--demixed transition
can be computed via Pirogov--Sinai theory~\cite{PS}
yielding,
\begin{equation}
M = z_d + 2 - 2/ 3 z_d^2 + \cdots
\end{equation}
which matches up smoothly with our MC results, see Fig.~3. 
It is easy to show that there is a 
demixing transition for $M \geq 2$~\cite{r3,r4,PS}, the
existence of sharp interfaces between coexisting phases, for $M=2$, on the
cubic lattice 
at large fugacity $z$ was proven in ~\cite{r4a}.

We next present results of our simulations and refer the reader to~\cite{LMNS}
for a more detailed description of the model and additional references.

\section{Results}

Our MC simulations were carried out on a 
bcc lattice of size $2 \times S^3 = 2 \times 22^3$
with periodic boundary conditions.  On an initially empty lattice we
deposit particles chosen at random from the $M$ components at fugacity $z$
respecting the exclusion of different species occupying neighboring sites.
We then sequentially update the
lattice using a checkerboard algorithm resulting in a good
vectorization. An update of a lattice site $(i_1^{(s)},i_2^{(s)},i_3^{(s)})$
on one of the two simple cubic sublattices $s$, making up the bcc lattice, 
which is occupied by a particle of type $I$ ($I=0$ indicating an empty site)
is done as follows: We randomly choose a new trial particle of type $I_{tr}$, 
where $I_{tr}$ can have any 
integer value between $0$ and $M$ with equal probability. 
$I_{tr} = 0$ refers to an attempted removal 
of a particle $I \not= 0$ from the lattice site, 
which is successful, if a number $X$ randomly chosen with equal
probability between $0$ and $1$ is smaller than the inverse fugacity $1/z$.
When this occurs $I$ gets the value $0$, otherwise it remains unchanged,
$I_{tr} \neq 0$ refers to an attempted deposition of a particle of type
$I_{tr}$.  If $I=0$
then it is successful if each of the four nearest neighbor sites is either
empty or occupied by a particle of the same type ($I_{tr}$) and $X <
z$. In this case $I$ gets the value $I_{tr}$, otherwise it remains
unchanged. A direct replacement attempt of a particle $I\not=0$ surrounded
by eight empty nearest neighbor sites is always successful.  Typically in a
simulation run after an equilibration of $2\times 10^5$ Monte Carlo steps
(MCS) we update the lattice $10^5$ times, the
configuration of every tenth step is taken for the evaluation of the
averages. A typical run with $10^5$ MCS took about 1~CPU hours on
a CRAY--YMP.

Let $m(i_1^{(s)},i_2^{(s)},i_3^{(s)})$ denote the occupancy of a site, 
$m(i_1^{(s)},i_2^{(s)},i_3^{(s)}) = 0$ if
the site $(i_1^{(s)},i_2^{(s)},i_3^{(s)})$ is empty and
$m(i_1^{(s)},i_2^{(s)},i_3^{(s)})=1$ otherwise.
As observables we took histograms $P_L(\phi_c)$ 
of the order parameter $\phi_c$ for the crystal structure
and $P_L(\phi_d)$ of the order parameter $\phi_d$ for
the demixed phase in subsystems of size $ 2 \times L^3$,
\begin{equation}
\phi_c = \frac{1}{2L^3} \sum_{s=1}^2 \sum_{i_1^{(s)},i_2^{(s)},i_3^{(s)} = 1}^L 
\left[2 m(i_1^{(s)},i_2^{(s)},i_3^{(s)})-1 \right](-1)^{2*s-1}
\label{pc}
\end{equation}
and
\begin{equation}
\phi_d = \frac{1}{2L^3} {\rm Max}_I N_L(I) -\rho/M
\label{pd}
\end{equation}
where 
$N_L (I)$ denotes the number of particles of type $I$ in a subsystem of
size $2 \times L^3$ and $\rho$ is the average overall density.

\subsection{Gas--Crystal Phase Transitions}

For a given $M$ the transition activity $z_c$ is found
by finite size scaling techniques~\cite{PRIV,BIND1}. 
In particular the $k-th$ moments of the order parameter distribution
$P_L(\phi_c)$,
\begin{equation}
< \phi_c^k >_L:= \int \phi_c^k P_L (\phi_c) d\phi_c
\label{cmoments}
\end{equation}
can be evaluated in subsystems of size $2 \times L^3$, and from them the
fourth order cumulant~\cite{BIND1} $U_L$\ \ ,

\begin{equation}
U_L = 1 - \frac{<\phi_c^4>_L}{3< \phi_c^2 >_L^2}\ .
\label{ccumu}
\end{equation}
In a one phase region far away from a critical point the subsystem size
typically can be chosen larger than the correlation length $\xi$, $L >>
\xi$ and the order parameter distribution is to a good approximation a
Gaussian centered around $0$, resulting in $U_L \to 0$ for $L \to \infty$.
In the two phase coexistence region far away from a critical point we can
again assume $L >> \xi$ and the order parameter distribution is bimodal
resulting in $U_L \to 2/3$ for $L \to \infty$.  Near the critical point
however we have $ L << \xi$, and using scaling arguments~\cite{BIND1} the
cumulant is a function of $L/ \xi$, resulting for $\xi \to \infty$ in the
same value of $U_\ast$ for all different $L$.  This method allows the
efficient determination of critical points by analyzing the cumulants for
different values of $z$ on different length scales $L$. Applied to our
model we should see,  for low values of
$z$, when the system is in the disordered one phase region, $U_{L'} > U_L$ for $L'
< L$.  Increasing $z$, we obtain, for large enough $M$,  a crystal phase with $U_{L'} < U_L$
for $L' < L$.  Near $z_c$ we  expect $U_{L'} \approx U_L$
for $L' \neq L$. This method for locating the transition fugacities
was used in our previous studies~\cite{LNS,LMNS} on the square lattice as well.

For $M= 3, 4, 5, 10$ we obtain in this way values of 
$z_c = 0.525 \pm 0.025, 0.3, 0.21, 0.085 \pm 0.01$ 
respectively, for $M=15, 17, 19$, 
$z_c = 0.055, 0.047,0.04 \pm 0.001$
and for $M=1000$, $z_c = 0.000725 \pm 0.000025$.
Cumulants for the gas--crystal transition of the hard
diamond and the $M=3$--component system are shown in Figs.~1 and 2,
respectively.
The transition points are presented in Fig.~3 together with
the asymptotic expression $Mz_c = 0.71$.
This corresponds to the value of $y_c$ on the bcc lattice
$y_c = 0.71 \pm 0.01$ which we obtained using MC techniques.
We also show in Fig.~3 a purely empirical fit of the $1/M^2$--corrections
for $z_c$, $z_c=0.71/M + C/M^2$, to the MC data, with a value of $C=2 \pm 0.5$.
The minimum number of components required for the existence of the crystal 
phase on the bcc lattice is $M = 3$.
In Fig.~4 we show the approach of the critical fugacity $Mz_c$ to
the limit of the hard diamond system.

\subsection{The Demixing Transition}

For $M=2$ we obtain a direct transition from the gas phase to the demixed phase
at a transition fugacity of $z=0.55 \pm 0.02$; the transition 
is second order with Ising exponents as on
a simple cubic lattice studied by Dickman et al.~\cite{Dick}.

For $M>2$ we observe a direct first order transition from the
crystal to the demixed phase.  This transition was analyzed by studying 
the order parameter distribution $P_L(\phi_d)$.  In the simulations we
find a hysteresis region around $z_d$, going approximately between 
two values of $z$, say $z_1$ and $z_2$, when increasing and decreasing the
fugacity. In
cases of a small hysteresis region with extent of less than $|z_1-z_2| <
0.1$, $z_d$ was taken as the average of this region, $z_d = (z_1+z_2)/2$. 
In cases of a larger hysteresis region we located the
transition fugacity $z_d$ by finding the 
relative stability of one of the two phases during a
simulation starting from configurations with both phases present in
parallel slices extending over the length of the simulation box,
as described in~\cite{LMNS}.
The resulting phase transition values of $z_d$ are shown in Fig.~1.
We note that with increasing number of
components, the transition fugacities approach the exact asymptotic line
$M=z+2-2/3z^2 + \cdots$.

\acknowledgements 
We thank A. Mazel for very helpful discussions.  
P.N. acknowledges support from the Deutsche
Forschungsgemeinschaft (Heisenberg foundation), the computations were
carried out at the CRAY-YMP of the RHRK Kaiserslautern and the CRAY-T90
of the HLRZ J\"ulich.  J.L.L. was supported by NSF Grant DMR 95-23266 and
DIMACS and its supporting agencies, the NSF under contract STC-91-19999 and
the N.J. Commission on Science and Technology.

%%%%%%%%%%%%%%%%%%%%%%%%%%%%%%%%%%%%%%%%%%%%%%%%%%%%%%%%%%%%%%%%%%%%%%%%%%%
%%%%%%%                       Appendices
%%%%%%%%%%%%%%%%%%%%%%%%%%%%%%%%%%%%%%%%%%%%%%%%%%%%%%%%%%%%%%%%%%%%%%%%%%%

%%\appendix
%%\section{}
%%\label{app1}

%%%%%%%%%%%%%%%%%%%%%%%%%%%%%%%%%%%%%%%%%%%%%%%%%%%%%%%%%%%%%%%%%%%%%%%%%%%
%%%%%%%%                       FIGURES
%%%%%%%%%%%%%%%%%%%%%%%%%%%%%%%%%%%%%%%%%%%%%%%%%%%%%%%%%%%%%%%%%%%%%%%%%%%

\unitlength1mm
\begin{figure}[hbt]
%\begin{center}
%\begin{picture}(70,70)
%\put(0,0){\psfig{figure=wrbcc.ps,width=70mm,height=70mm}}
%\end{picture}
\caption[]{
Cumulants versus $z$ for the gas--crystal transition for $M=3$.
Symbols indicate different subsystem sizes, lines are for visual help.
}
%\end{center}
\label{FIG1}
\end{figure}

\begin{figure}[hbt]
%\begin{center}
%\begin{picture}(70,70)
%\put(0,0){\psfig{figure=wrbcc.ps,width=70mm,height=70mm}}
%\end{picture}
\caption[]{
Cumulants versus $z$ for the gas--crystal transition for the
hard diamond system on the bcc lattice.
Symbols indicate different subsystem sizes, lines are for visual help.
}
%\end{center}
\label{FIG2}
\end{figure}

\begin{figure}[hbt]
%\begin{center}
%\begin{picture}(70,70)
%\put(0,0){\psfig{figure=wrbcc.ps,width=70mm,height=70mm}}

%\end{picture}
\caption[]{
Phase diagram in the $M$--$z$ plane for a bcc lattice (MC).
Full lines:  Asymptotic lines for the phase transitions in the high fugacity
region, $M=z+2-2/3z^2$, and for the transition in the low fugacity region,
$Mz=0.71 \pm 0.01$, dashed line: transition line in the low fugacity region
by fit of the $1/M^2$ corrections for $z_c$ to the MC data, 
$z_c = 0.71/M + C/M^2$, with $C = 2 \pm 0.5$.
Symbols for MC: Transition points from the
gas phase to the crystal phase (circles), from the gas to the demixed
phase (squares) and from the crystal to the demixed phase (triangles).
}
%\end{center}
\label{FIG3}
\end{figure}

\begin{figure}[hbt]
%\begin{center}
%\begin{picture}(70,70)
%\put(0,0){\psfig{figure=wrbcc.ps,width=70mm,height=70mm}}
%\end{picture}
\caption[]{
Critical fugacities $Mz_c$ versus $1/M$ for the gas--crystal transition.
The open symbols are results of the present Monte Carlo simulations,
The black symbol (with error bars) on the $y$--axis indicates the
critical fugacity of the hard diamond system from series
expansions~\cite{GAUNT},
the open symbol on the $y$-axis the critical fugacity from our
MC simulations.
Lines are for visual help.
}
%\end{center}
\label{FIG4}
\end{figure}

%%%%%%%%%%%%%%%%%%%%%%%%%%%%%%%%%%%%%%%%%%%%%%%%%%%%%%%%%%%%%%%%%%%%%%%%%%%
%%%%%%%                       TABLES
%%%%%%%%%%%%%%%%%%%%%%%%%%%%%%%%%%%%%%%%%%%%%%%%%%%%%%%%%%%%%%%%%%%%%%%%%%%

%
% Here is an example of the general form of a table:
% Fill in the caption in the braces of the \caption{} command. Put the label
% that you will use with \ref{} command in the braces of the \label{} command.
% Insert the column specifiers (l, r, c, d, etc.) in the empty braces of the
% \begin{tabular}{} command.
%
% \begin{table}
% \caption{}
% \label{}
% \begin{tabular}{}
% \end{tabular}
% \end{table}
%
%\end{multicols}
\end{document}